# Proton and Li-ion permeation through graphene with eight-atom-ring defects

E. Griffin[1], L. Mogg[1], G. P. Hao[1,2], K. Gopinadhan[1,3], C. Bacaksiz[4], G. Lopez-Polin[1,5], T. Y. Zhou[6], V. Guarochico[1], J. Cai[1], C. Neumann[7], A. Winter[7], M. Mohn[8], J. H. Lee[9], J. Lin[10,11], U. Kaiser[8], I.V. Grigorieva[1], K. Suenaga[10], B. Özyilmaz[9], H. M. Cheng[6,12], W. C. Ren[6], A. Turchanin[7], F. M. Peeters[4], A. K. Geim[1], M. Lozada-Hidalgo[1]

[1]Department of Physics and Astronomy & National Graphene Institute, The University of Manchester, Manchester M13 9PL, UK
[2]State Key Laboratory of Fine Chemicals, School of Chemical Engineering, Dalian University of Technology, Dalian 116024, China
[3]Department of Physics, Indian Institute of Technology Gandhinagar, Gujarat 382355, India
[4]Departement Fysica, Universiteit Antwerpen, Groenenborgerlaan 171, B-2020 Antwerp, Belgium
[5]Departamento de Física de la Materia Condensada, Universidad Autónoma de Madrid, 28049 Madrid, Spain
[6]Shenyang National Laboratory for Materials Science, Institute of Metal Research, Chinese Academy of Sciences, Shenyang 110016, China
[7]Institute of Physical Chemistry, Friedrich Schiller University Jena, 07743 Jena, Germany
[8]Central Facility for Electron Microscopy, Electron Microscopy Group of Materials Science, Ulm University, Ulm 89081, Germany
[9]Department of Physics, Department of Materials Science and Engineering & Centre for Advanced 2D Materials, National University of Singapore, Singapore
[10]National Institute of Advanced Industrial Science and Technology, Tsukuba, Japan & Department of Mechanical Engineering, The University of Tokyo, Japan
[11]Department of Physics, Southern University of Science and Technology, Shenzhen 518055, China
[12]Shenzhen Graphene Center, Tsinghua-Berkeley Shenzhen Institute, Tsinghua University, Shenzhen 518055, China

**Defect-free graphene is impermeable to gases and liquids[1–4] but highly permeable to thermal protons[5–8]. Atomic-scale defects such as vacancies, grain boundaries and Stone-Wales defects are predicted[9–11] to enhance graphene's proton permeability and may even allow small ions through, whereas larger species such as gas molecules should remain blocked. These expectations have so far remained untested in experiment. Here we show that atomically thin carbon films with a high density of atomic-scale defects continue blocking all molecular transport, but their proton permeability becomes ~1,000 times higher than that of defect-free graphene. Lithium ions can also permeate through such disordered graphene. The enhanced proton and ion permeability is attributed to a high density of 8-carbon-atom rings. The latter pose approximately twice lower energy barriers for incoming protons compared to the 6-atom rings of graphene and a relatively low barrier of ~0.6 eV for Li ions. Our findings suggest that disordered graphene could be of interest as membranes and protective barriers in various Li-ion and hydrogen technologies.**

Despite being a one-atom-thick material, no more than a few gas atoms per hour can permeate through micrometer-sized defect-free graphene membranes, as proven experimentally[3]. Even the smallest ions are blocked by the crystal[4]. These phenomena arise because the dense electron clouds of graphene's crystal lattice impose energy barriers of several eV to incoming molecular and ionic species[9–11], which forbids their permeation under ambient conditions. In contrast, it has been shown experimentally that protons, nuclei of hydrogen atoms, can transport through defect-free graphene relatively easily, overcoming an energy barrier of only ≲1 eV (refs [3–6]). In this context, theory predicts that modifying graphene's lattice by introducing 7- or 8- atom rings should greatly reduce the energy barriers faced by protons[9] and may even allow small ions (e.g. $Li^+$)[10] to permeate. This is without losing graphene's impermeability with respect to atoms and molecules[11]. However, the permeability of such 'extended' carbon rings remains untested experimentally, mostly because large enough graphene samples with high-density of atomic-scale defects remained elusive. To create defects, graphene crystals were previously perforated using ion irradiation or chemical and plasma etching[12–14]. This approach results in a local loss of carbon atoms[12–14] that typically form nanometer-sized pores[12] rather than atomic-scale defects. These nanopores are permeable to gases[2,12], ions[12] and even macromolecules (e.g., DNA)[12,15]. An alternative approach is to grow materials with the required extended carbon-atom rings from the outset. Recently, high-quality one-atom-thick films with a high density of these 8-atom defects have been demonstrated using laser-assisted chemical vapor deposition[16] and high-temperature quenching of metal foils in liquid hydrocarbon compounds[17]. We refer to the latter two materials as disordered graphene (DG). Unlike irradiated graphene, DG presents a dense net of the 'extended' carbon rings over the entire area, which allows for permeability studies of these ring structures. Below, we report proton and lithium-ion transport through disordered graphene grown using the methods reported in refs. [16] and [17].

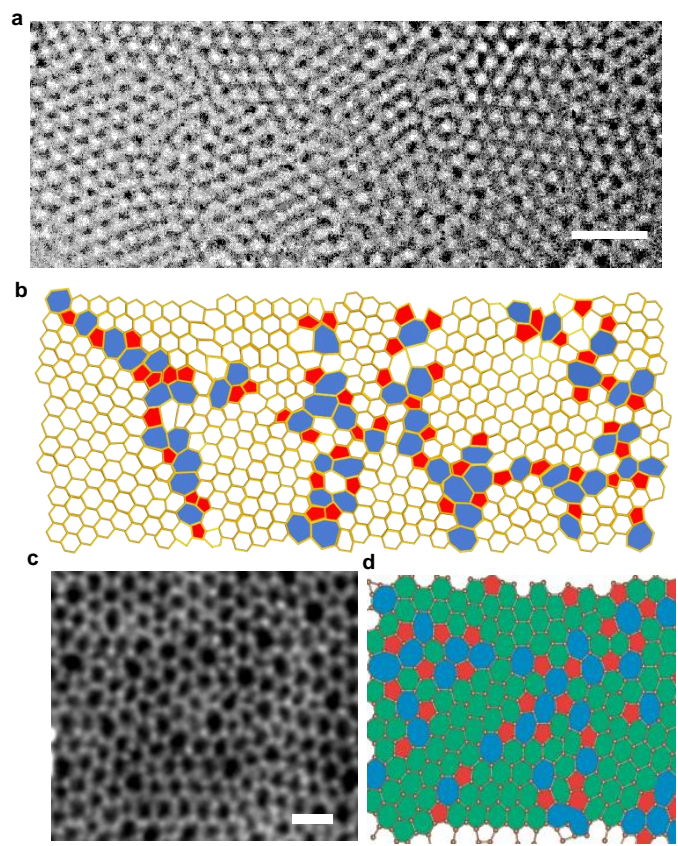

**Figure 1 | Transmission electron microscopy characterization of disordered graphene. a,** High resolution TEM micrograph of nanocrystalline graphene. Scale bar, 1 nm. **b,** Schematic of the image in panel (a). 5-atom ring structures are marked in red; 6-atom rings in white with yellow boundaries; both 7- and 8- atom rings in blue. **c,** High-angle annular dark-field scanning TEM image of monolayer amorphous carbon. Scale bar, 0.5 nm. **d,** Schematic of the image in (c). Red and green areas indicate 5- and 6- atom rings, respectively. The blue-colored areas denote 7- and 8- atom ring structures.

To characterize the DG materials with atomic precision, we used transmission electron microscopy (TEM). The two-dimensional (2D) materials were suspended over circular apertures (~0.1 μm in diameter) etched in free-standing silicon nitride (SiN) membranes[5] (Supplementary Figure 1). Figs 1a,b show that DG grown by quenching in liquid hydrocarbons[17] can be described as a patchwork of nanometer-sized graphene crystallites. The grain boundaries of the material consist of 5-, 6-, 7- and 8-atom rings. Because of the small domain size (3-4 nm), the material contains a large density of these non-hexagonal structures. We refer to this material as nanocrystalline graphene. On the other hand, DG synthesized via laser assisted chemical vapor deposition[16] consists of a one-atom-thick amorphous assembly of 5-, 6-, 7- and 8-carbon-atom rings (Figs 1c,d), without any visible presence of graphene crystallites. We refer to the material as monolayer amorphous carbon (MAC)[16]. The two DG films were compared with two other reference materials. The first one was an amorphous nanometer-thick carbon film referred to as carbon nanomembrane (CNM). This material was synthesized by self-assembly of aromatic precursors that were cross-linked by electron irradiation, which resulted in short-range-order molecular nanosheets[18,19] and a dense (~$10^{14}$ cm$^{-2}$) network of sub-nanometer (~0.7 nm in diameter) pores piercing CNMs[19]. In this work, we focused on ~0.9 and 1.2 nm thick films from 1,1'-biphenyl-4-thiol and [1'',4',1',1]-terphenyl-4-thiol precursors, known as BPT-CNM and TPT-CNM, respectively[18] (Supplementary Figure 2). The second reference material was defect-free graphene crystals obtained by mechanical exfoliation. Accordingly, our study compares the permeability of 2D carbon materials over the entire range of their possible disorder, from crystalline to disordered, to amorphous structures.

To further characterize the DG materials, we studied their mechanical properties. The suspended membranes were indented at their center with an atomic force microscope (AFM) tip and the deflection, $\delta$, of the membrane was recorded as a function of applied force, $F$. From the measured $F(\delta)$, it is possible to extract the 2D elastic (Young) modulus[20]. It was found ~190 and 100 N m$^{-1}$ for nanocrystalline graphene and amorphous carbon, respectively (Supplementary Figure 3). The reference graphene and CNM exhibited moduli of ~340 and 10 N m$^{-1}$, respectively, in agreement with the previous measurements[20,21]. This shows that the rigidity of 2D carbon decreases with increasing disorder. Similarly, the mechanical strength also decreased by a factor of 6 for amorphous carbon compared to defect-free graphene ('Mechanical properties of disordered graphene' in Supplementary Information). These changes are attributed to the presence of 7- and 8- atom rings, which weakens disordered graphene. Nevertheless, DG remains 10 times stronger than nanometer-thick carbon membranes. This can be attributed to the fact that DG is formed mainly by strong $sp^2$ carbon-carbon bonds[16,17], unlike CNM[18].

To find out whether the studied materials allow molecular transport, we measured their permeability with respect to helium, the most permeable gas. Before any measurements, our membranes were studied using atomic force and scanning electron microscopy. Membranes with cracks or other visible imperfections were discarded, and only the rest were tested for helium permeation ('Device fabrication and characterization' in Supplementary Information). They separated two chambers. The feed chamber had a helium gas at a controllable pressure, and the permeate chamber was evacuated and connected to a mass spectrometer. Typically, the feed pressure was slowly increased up to a few tens of mbar to avoid damaging the membranes. No helium flow could be detected through either DG or CNM membranes within the accuracy of the mass spectrometer, which sets an upper bound for their helium permeance of ~$10^{-14}$ mol s$^{-1}$ cm$^{-2}$ Pa$^{-1}$. This limit is comparable to, or even lower than, that of commercially-available ion conductive polymer membranes of >100 μm in thickness, which are optimized to block gas permeation[22] ('Gas transport measurements' in Supplementary Information).

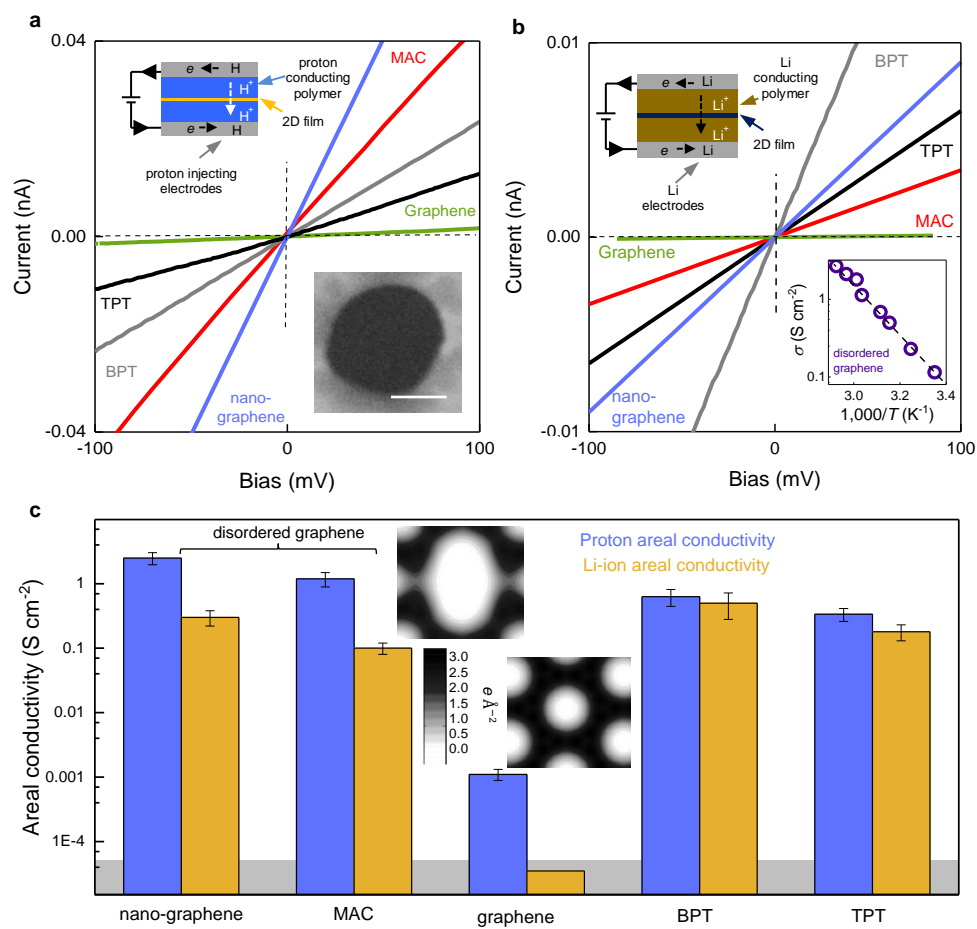

**Figure 2 | Proton and Li-ion transport. a**, Examples of *I-V* characteristics for proton transport through different carbon films (color coded). Nano-graphene stands for nanocrystalline graphene. Top inset, schematic of the experimental setup. Bottom inset, scanning electron micrograph of a suspended MAC membrane. The dark circular area in the image corresponds to the aperture in the SiN substrate over which the 2D film was suspended. Scale bar, 100 nm. **b**, Examples of *I-V* characteristics found in Li-ion transport measurements. Top inset, device schematic. Bottom inset, Arrhenius plot for a typical DG membrane. Dotted line, guide to the eye. **c**, Statistics for proton (blue) and Li-ion (brown) areal conductivities measured for different 2D carbon films. Each bar shows the average for at least three different devices. Error bars, standard error of mean. The grey area indicates our detection limit given by parasitic leakage currents. Insets, charge density integrated along the direction perpendicular the graphene plane for a 5-8-5 defect (top) and defect-free graphene (bottom). The white areas represent minima in the electron density at the center of the ring structures.

These helium-tight carbon films should in principle be more permeable to protons than defect-free graphene because of their 'looser' structures. To investigate proton permeation, the suspended membrane devices described above were coated on both sides with a proton conducting polymer (Nafion[23]) and electrically contacted with two proton injecting electrodes (inset of Fig. 2a and Supplementary Figure 4), following the recipe reported previously[5]. In this setup, the films effectively act as barriers between two semi-infinite proton reservoirs. To perform electrical measurements, the devices were placed in a humid $H_2$ atmosphere which ensured the high proton conductivity of Nafion[5] (see 'Electrical measurements' in Supplementary Information). The current density *I* was found to vary linearly with small applied voltage *V* (Fig. 2a), which allowed us to determine their areal conductivity, $\sigma = I/V$. We found that all the disordered carbon films were ~1,000 times more permeable than defect-free graphene[5]. The most proton conductive was nanocrystalline graphene, closely followed (factor of ~2 lower) by MAC. Carbon nanomembranes (BPT-CNM and TPT-CNM) exhibited $\sigma$ lower than MAC and $\sigma_{BPT}^{H} \sim 2\,\sigma_{TPT}^{H}$, which means their conductivities approximately scaled with their thicknesses. For reference, we measured devices with no carbon films placed over the apertures. Their resistance was ~100 times smaller than that of any device with the tested 2D carbon, which ensured that the resistance stemming from Nafion had a negligible contribution into the measured $\sigma$.

The high proton permeability of the disordered carbon films suggests that they could also be permeable to other ions. To investigate this possibility, we used Li ions, the penultimate smallest ion after the proton. In lithium transport experiments, suspended membranes were coated on both sides with a Li-conducting polymer[24] and electrically connected to Li metal electrodes (see 'Electrical measurements' in Supplementary Information). During assembly and electrical measurements, the devices were kept inside a glovebox containing an inert gas atmosphere (less than 1 ppm of either water or oxygen) to prevent Li from reacting. Fig. 2b shows that *I-V* characteristics of all Li-transport devices were linear. All the 2D membranes were less permeable to Li ions than protons. BPT-CNM and TPT-CNM were the most Li-ion conductive of the tested materials displaying $\sigma$ rather close to the values found for proton transport, which means little selectivity between protons and Li ions. Again, their Li-ion conductance roughly scaled with their thickness. In contrast, the Li-ion conductivities for both DG materials were ~10 times smaller than the corresponding values found for proton transport – that is, disordered graphene displayed large proton/Li-ion selectivity. As for defect-free graphene, we could not discern any Li-ion transport through it, even if we increased the membrane areas by two orders of magnitude. Our sensitivity level of ~$10^{-13}$ A, given by leakage currents, translates the latter findings into an upper bound for Li-ion transport through defect-free graphene of ~$10^{-5}$ S cm$^{-2}$. As in the case of proton transport measurements, the lithium polymer was sufficiently conductive to contribute little into the reported $\sigma$ ('Electrical measurements in Supplementary Information).

Our results suggest that the energy barriers for proton and Li-ion transport through disordered graphene should be lower than those of defect-free graphene. To quantify the barriers, we measured the temperature (*T*) dependence of $\sigma$ between ~2 and 50 °C. This *T*-range ensured adequate performance of the ion conducting polymers[23,24] and high reproducibility of $\sigma(T)$ between different devices and for consecutive heating and cooling cycles (Supplementary Figure 5). We found that Li-ion transport for both DG materials increased rapidly with *T* and could be described by the Arrhenius relation $\sigma \propto \exp(-E/kT)$ with the same activation energy $E = 0.62\pm0.06$ eV where $k$ is the Boltzmann constant (Fig. 2b, inset). For proton transport, $\sigma(T)$ was less reproducible between different devices at high *T*, and we had to limit our analysis to temperatures below room-*T*. We attribute this to the fact that elevated *T* are known to cause moisture loss in Nafion, which reduces its conductivity[23] and introduces a non-negligible series resistance. Even within the limited *T*-range, we found that $\sigma$ increased notably with *T*, and could be fitted by $E \approx 0.4$ eV for both DG materials (Supplementary Figure 5). Note that, even at our highest *T* of 50 °C, Li-ion transport could not be detected through defect-free graphene. This is perhaps not surprising given the large activation energy of ~0.8 eV faced by small-size protons[5]. A much higher barrier can be expected for larger Li ions[10], effectively forbidding Li-ion transport to be detectable.

The enhancement of proton and Li-ion permeability in DG with respect to defect-free graphene can be qualitatively understood from the perspective of electron clouds, which present transport barriers[5]. The inset of Fig. 2c shows that the electron clouds surrounding 8-atom ring structures are notably sparser than those around 6-atom rings. This should make the former rings more permeable to ions, as shown by similar analysis for graphene and hexagonal boron nitride[5]. This interpretation is substantiated by density functional theory (DFT) calculations. We find that the energy barrier for both proton and Li-ion penetration through carbon-ring structures decreases with increasing number of atoms within the rings (Supplementary Fig. 6), in agreement with previous theory results[9,10]. Because of the exponential dependence of $\sigma$ on $E$, proton and Li-ion transport through disordered graphene should probably be dominated by contributions from 8-atom rings, even if these were relatively rare[16,17]. Our DFT calculations also explain why protons permeate ~10 times faster than Li ions through disordered graphene (Fig. 2c). 8-atom rings provide an energy barrier for incoming Li ions approximately twice higher than for protons. This interpretation is also consistent with the absence of proton/Li-ion selectivity in carbon nanomembranes, which do not have 8-atom ring structures. Indeed, $\sigma$ measured for CNM scaled with the thickness for both H and Li ions, indicating that the bulk transport is important, rather than the single, entry-exit barrier presented by one-atom-thick graphene materials. The conclusion about bulk transport through CNMs is also consistent with their microscopic structures (effectively a dense network of sub-nanometer pores[19]) such that ions diffuse along tortuous trajectories as typical for porous media[25].

Besides providing fundamental insights into ion transport through 2D materials, disordered graphene is interesting in terms of applications. The DG materials reach technologically relevant proton conductivities at notably lower temperatures than defect-free graphene. Our experimental data yield that the proton areal conductivity of disordered graphene at ~60 °C should exceed the industry benchmark set by Nafion 117[26] (~5 S cm$^{-2}$). Defect-free graphene reaches this level only at ~200 °C. However, it is this latter temperature that is most desirable for fuel-cell operation[27–29]. By extrapolation of the measured $\sigma(T)$, our DG membranes can reach ~100 S cm$^{-2}$ for this $T$ range, well above the industry targets[27]. Importantly, disordered graphene can be mass produced[16,17] and fabricating large-area proton conducting membranes should also be straightforward, as demonstrated for the case of defect-free graphene[30]. Furthermore, the lithium permeability of DG deserves special attention. Graphene is being explored as a material to host highly reactive Li-metal[31] and Li-Si particles[32] as anodes and Li-sulfur as cathodes[33] in batteries, in order to protect them from chemical reactions with electrolytes, prohibit Li dendritic growth and provide mechanical stability. The key properties needed for the latter applications are high Li-ion conductivity combined with impermeability to reactive species. Because defect-free graphene is impermeable to Li-ions, defects are essential, and those permeable only to Li ions (like 8-atom rings) offer considerable advantages. All the above indicates that defect engineering in graphene and other 2D materials could be a productive venue for optimizing their use in energy conversion and storage technologies.

# Proton and Li-ion permeation through graphene with eight-atom-ring defects

E. Griffin, L. Mogg, G. P. Hao, K. Gopinadhan, C. Bacaksiz, G. Lopez-Polin, T. Y. Zhou, V. Guarochico, J. Cai, C. Neumann, A. Winter, M. Mohn, J. H. Lee, J. Lin, U. Kaiser, I.V. Grigorieva, K. Suenaga, B. Özyilmaz, H. M. Cheng, W. C. Ren, A. Turchanin, F. M. Peeters, A. K. Geim, M. Lozada-Hidalgo

**Materials synthesis**

Nanocrystalline graphene films were synthesized by quenching a Pt foil in liquid ethanol, as described in a recent report[1]. Pt foil (99.95 wt%, 150 μm thickness) was finely polished and annealed at 800 °C in air for 1 h as a cleaning step. After heating at 900 °C in an Argon atmosphere, the Pt foil was rapidly quenched in ethanol at room temperature to grow a large-area nanocrystalline graphene (NG) film. See ref. [1] for further details. These films were then transferred on silicon-oxide substrates using the electrochemical bubbling method[2]. In brief, poly-methyl-methracrylate (PMMA) was spin-coated on a NG film with Pt foil substrate. The PMMA-coated NG film/Pt foil was immersed into a NaOH (1 M) aqueous solution and used as a cathode. A Pt wire was used as anode and a constant electric current of 0.2 A was applied between the two electrodes. After the PMMA-coated NG film separated from the Pt substrate by hydrogen bubbles, it was cleaned in deionized water several times and collected onto a silicon-oxide substrate. The PMMA was then dissolved in acetone and isopropyl alcohol.

Monolayer amorphous carbon (MAC) films were synthesized by laser-assisted chemical vapor deposition (LCVD), as described in a recent report[3]. Cu foils (35 μm thick) were cleaned and annealed in a hydrogen atmosphere at 1010 ˚C. The foil was placed in the LCVD chamber, which was evacuated and then filled with methane gas. A plasma (350 kHz pulsed DC generator at 5 W) was turned on away from the sample and the substrate was directly exposed to a pulsed krypton fluoride laser (40-75 mJ cm$^{-2}$, 50 Hz). This process yields MAC films on both faces of the Cu foil. MAC films were then transferred onto silicon-oxide substrates. To that end, one of the faces of the Cu foil was spin-coated with PMMA and the other face was exposed to oxygen plasma to remove the MAC film on that surface of the foil. The PMMA-coated MAC film/Cu foil was then placed in ammonium persulfate solution to dissolve the Cu foil. After cleaning with deionized water several times, the MAC film was collected onto a silicon-oxide substrate and the PMMA was dissolved.

Carbon nanomembranes were synthesized by self-assembly of molecular precursors[4]. Mica backed Au layers were placed in solutions of either 1'-biphenyl-4-thiol or [1'',4',1',1]-terphenyl-4-thiol precursors. The precursor materials self-assemble on the Au surface forming a continuous layer, due to the van der Waals interactions between the carbon atoms in the molecules. These self-assembled monolayers are then cross-linked by electron irradiation. The resulting films are spin coated with PMMA. The mica substrate is then mechanically removed from the back of the Au foil, leaving the PMMA coated- carbon film/Au layer. The Au film is then dissolved in Au-iodide solution. After several steps of cleaning in deionized water, the disordered carbon film was collected on a silicon-oxide substrate and the PMMA was dissolved.

**Device fabrication and characterization**

The device fabrication process starts by preparing substrates to suspend the 2D films. To that end, a silicon substrate coated on both sides with a 100 nm thick silicon-nitride layer is etched using a combination of chemical and reactive ion etching to produce a suspended silicon-nitride membrane (~50 μm wide). Then, using a combination of electron beam lithography and reactive ion etching, an aperture (0.1 μm radius) is etched in this membrane[5]. The next step is to examine the 2D films on silicon oxide substrates using an optical microscope. An area without cracks and visible imperfections is selected and suspended over the holes in silicon-nitride[5]. The resulting free-standing membrane is imaged using atomic force (AFM) and scanning electron microscopy (SEM). Membranes with holes or imperfections visible under AFM or SEM were discarded from the outset. Supplementary Figure 1 shows images of a typical membrane used in our gas and ion transport measurements. These membranes were continuous throughout and no imperfections were visible under AFM or SEM.

**Supplementary Figure 1 | Membrane characterization**. **a**, Atomic force micrograph of a suspended

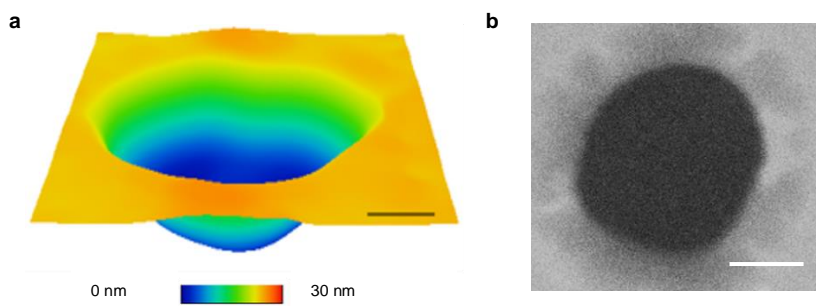

disordered graphene membrane. **b**, Scanning electron micrograph of the device in panel (a). Scale bars 100 nm.

**Transmission electron microscopy characterization**

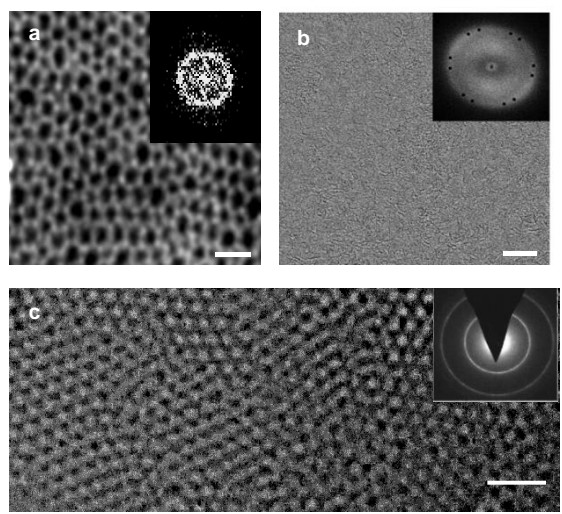

**Supplementary Figure 2 | Transmission electron microscopy characterization of 2D films**. **a**, STEM image of monolayer amorphous carbon. Inset, Fourier transform pattern. Scale bar 0.5 nm. **b**, HRTEM image of graphene-encapsulated TPT-CNM. Inset, Fourier transform pattern. The Fourier transform pattern of the two defect-free graphene layers used for encapsulation were filtered out. Scale bar 2 nm. **c**, HRTEM image of nanocrystalline graphene. Inset, diffraction pattern. Scale bar 2 nm. All Fourier transform diffraction patterns confirm that the films lack long-range order.

Disordered graphene samples for transmission electron microscopy characterization were prepared in the same way as those for ion transport, except no polymers or electrodes were attached to the sample. On the other hand, for carbon nanomembrane samples, the material was encapsulated with pristine graphene to protect it from electron irradiation and enhance contrast in the microscope. All samples were baked at ∼160 °C in a nitrogen atmosphere to reduce hydrocarbon contamination.

For HRTEM characterization, we used an FEI Titan microscope operating at a low electron beam voltage of 80 kV with an image spherical aberration corrector and a monochromator. Supplementary Figure 2 shows the Fourier transform and electron diffraction patterns of the samples. These confirm the absence of long-range order structure of the materials. For carbon nanomembranes, the Fourier transform patterns of the pristine graphene layers used to encapsulate the sample were filtered out. For STEM characterization of monolayer amorphous carbon, we used a JEOL 2100F with a delta probe aberration corrector. The acceleration voltage used was 60 kV, the convergent illumination angle was 35 mrad and the detector angle ranged from 45-200 mrad. For STEM images, the samples were heated to 700 ºC using the heating-holder in the microscope.

**Mechanical properties of disordered graphene**

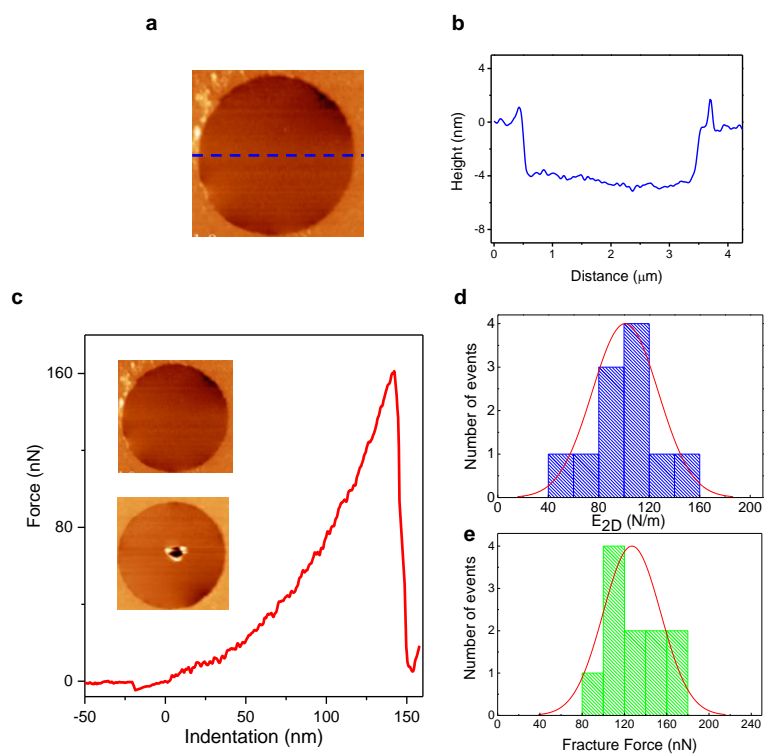

**Supplementary Figure 3 | Mechanical properties of monolayer amorphous carbon. a**, AFM image of a suspended monolayer amorphous carbon (MAC) membrane. **b**, Height profile of membrane in panel (a) in the absence of any applied force. **c**, $F(\delta)$ curves for a typical MAC membrane. At large indentation (~150 nm) the membrane breaks. Top inset shows AFM images of the membrane before and after being punctured by the AFM tip. **d**, Statistics of two dimensional elastic modulus. **e**, Statistics of fracture force.

To study the mechanical properties of disordered graphene, we suspended the films over circular apertures of radius $a \approx 1.5$ μm etched in silicon oxide substrates, as previously reported[6]. Qualitatively, the mechanical robustness of the materials is demonstrated by the fact that freestanding membranes can be fabricated in the first place. However, to quantitatively characterize their stiffness and strength, the suspended films were indented at their center with an atomic force microscope (AFM) tip and the deflection, $\delta$, of the membrane was recorded as a function of applied force, $F$. The resulting indentation curves for monolayer amorphous carbon films are shown in Supplementary Figure 3. Similar curves were obtained for nanocrystalline graphene, as previously reported[1]. These can be approximated by the equation: $F(\delta) = \pi\sigma_0\delta + q^3(E_{2D}/a^2)\delta^3$ where $\sigma_0$ is the membrane pre-tension, $q$ is a dimensionless constant and $E_{2D}$ is the two-dimensional elastic modulus[6]. From the measured $F(\delta)$ it is possible to extract $E_{2D}$ =100 ±30 N m$^{-1}$ for disordered-graphene. By further increasing the force applied to the suspended membrane, it is possible to puncture it. From the maximum force that the membrane can withstand ($F_{max}$) it is possible to obtain the ultimate strength by $\sigma_{2D}^{max} = \sqrt{\frac{F_{max}E_{2D}}{4\pi R_{tip}}}$, with $R_{tip}$ the radius of the tip. This yields to $\sigma_{2D}^{max}$ = 5 ±0.5 N m$^{-1}$. In comparison, the ultimate strength of graphene is ≈30 N m$^{-1}$ [6].

**Electrical measurements**

Suspended membrane devices fabricated as described above were coated with ion conducting polymers and ion injecting electrodes. For proton transport devices, the membranes were coated with Nafion solution (5%, 1100 equivalent weight) on both sides and then electrically connected with porous carbon electrodes containing Pt catalyst (20% Pt on carbon). The devices were then baked at 130 ˚C in a humid atmosphere to cross-link the polymer, as described in a previous report[5]. For Li-ion transport, the suspended membrane devices were placed inside a glovebox containing less than 0.5 parts-per-million of both water and oxygen. These were then coated on both sides with a standard lithium conducting polymer, as described in previous reports of Li-ion studies in 2D materials systems[7,8]. The polymer used consisted of LiTPFSI salt[7,9] dissolved in polyethylene oxide (PEO). To prepare the polymer, 0.05 g of LiTFSI and 0.3 g of powdered PEO (100,000 molecular weight) were dried overnight at 180 ˚C and 60 ˚C, respectively. These were then mixed in a 38:1 molar ratio with 2 ml of acetonitrile (99.8% anhydrous and further dried with 3Å molecular sieves) and left stirring overnight at room temperature inside a glovebox, as described in a previous report. Devices were then electrically connected with Li electrodes. See Supplementary Figure 4.

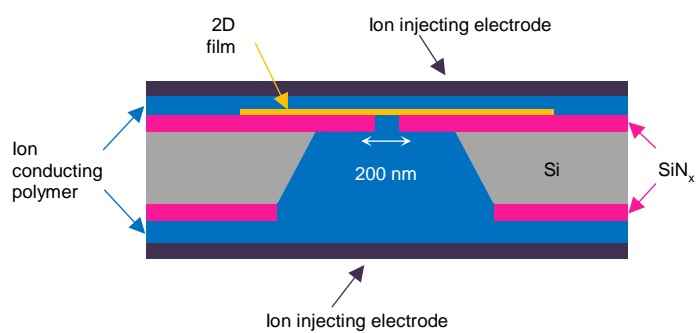

**Supplementary Figure 4 | Schematic of ion transport devices**. Schematic of suspended membrane devices assembled with ion conducting polymers and ion conducting electrodes.

Electrical measurements of Li-ion transport devices were conducted inside the same glovebox used for their assembly. Proton transport devices, on the other hand, were placed inside a chamber containing a 10% H$_2$ in Ar gas atmosphere at 100% relative humidity. To measure the *I-V* response of all devices, a Keithley SourceMeter 2636A was used to both apply voltage and measure current. Voltages were varied typically between ±200 mV using sweep rates of <0.1 V min$^{-1}$.

For reference, we measured devices without any carbon membrane covering the aperture in the silicon nitride substrates ('open hole devices'). For both Li-ion and Nafion polymers we found that open-hole devices displayed an areal conductivity of $\sigma_{open}$ ~ 100 S cm$^{-2}$. This value is ~100 times larger than that of devices with any of the membranes we measured for proton and Li-ion transport devices. This guarantees that the measured $\sigma$ in our devices with carbon membranes is dominated by ion transport through the 2D materials rather than by resistance from the polymers and contacts. Note that micro- and nano-scale devices allow for large current densities because of their size. The limiting conductance ($G$) of a cell with a small constriction of radius $r$ and electrolyte conductivity $\kappa$ is[5,10]: $G = 4\pi\kappa r$. In our case[4], $\kappa \sim 1$ mS cm$^{-1}$, $r \approx 100$ nm and hence we have $G \sim 0.1$ μS, which translates into $\sigma_{open} \sim 100$ S cm$^{-2}$ for the apertures in our devices, in agreement with the experiment.

**Gas transport measurements**

For gas transport measurements, suspended membrane devices (~1 μm diameter) without cracks or nanometer-sized imperfections, were clamped with O-rings and used to separate two chambers built from standard vacuum components. One of these chambers (permeate) was connected to a mass spectrometer (Inficon UL200) and the other (feed) was equipped with an electrically controlled dosing valve, which allowed us to slowly introduce helium into the chamber. Both chambers were initially evacuated to a pressure of ~$10^{-2}$ mbar. Then, using the dosing valve, we controllably introduced helium gas into the feed chamber until the pressure reached a few tens of mbar. We normally did not apply higher pressures as this could cause some of the 2D suspended membranes to break and leak. Within this pressure range, we could not detect any gas transport at all within the sensitivity of our mass spectrometer, $10^{-12}$ mbar l s$^{-1}$. The units from our mass spectrometer, mbar l s$^{-1}$, are straightforward to convert to mol s$^{-1}$ using the ideal gas law. For the applied pressure and membrane area, the found gas transport upper bound translates into an upper bound to the maximum gas flow possible through these membranes as ~$10^{-14}$ mol s$^{-1}$ cm$^{-2}$ Pa$^{-1}$. Note that this value is comparable to the gas permeability of ~180 μm thick Nafion membranes[11].

**Temperature dependence of areal conductivity**

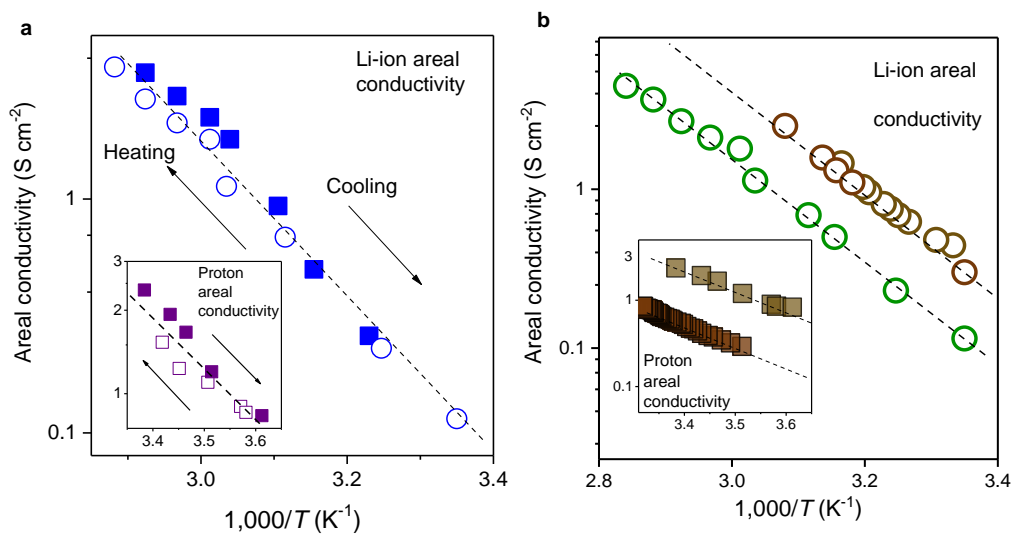

**Supplementary Figure 5 | Arrhenius plots of $\sigma$ of disordered graphene. a**, Arrhenius plots for heating (open symbols) and cooling (solid symbols) cycles of $\sigma(T)$ for Li-ion devices (main panel) and for proton transport devices (inset). Dotted lines, guide to the eye. **b**, Arrhenius plots of $\sigma(T)$ for different devices (marked with different colors) for lithium ion transport devices (main panel) and for proton transport devices (inset). Dotted lines, guide to the eye.

Supplementary Figure 5a shows that $\sigma(T)$ of both proton and Li-ion transport devices displayed negligible hysteresis between heating and cooling cycles. Furthermore, the $\sigma(T)$ data was also reproducible between different devices within our experimental accuracy of ~60 meV; in agreement with a previous report of ion transport through 2D materials[5] (Supplementary Figure 5b). To ensure this high reproducibility, we measured our devices well within the operating temperature of the ion conducting polymers. For Nafion devices, we normally measured from room-$T$ down to ~2 °C. This ensured that no dehydration of the Nafion polymer, with its concomitant conductivity reduction, took place. For Li-ion transport devices, we measured from room-$T$ up to a maximum of ~50 °C to avoid reaching the glass transition temperature of the polymer, as this could introduce mechanical stress that could damage our suspended one-atom-thick membranes.

**Density functional theory calculations**

Density functional theory was used to investigate the energy barriers to proton and Li ion permeation through disordered graphene. The calculations were performed using the projector augmented wave (PAW) method implemented in the Vienna *ab-initio* Simulation Package (VASP)[12–14]. To describe the electron exchange and correlation, the Perdew-Burke-Ernzerhof (PBE) form of the generalized gradient approximation (GGA) was adopted[15]. The van der Waals force, which is important for the layered materials, was taken into account by using DFT-D2 method of Grimme[16]. The calculations were performed using the following parameters. The kinetic energy cutoff of the plane-wave basis set was 500 eV in all calculations. The total energy difference between the sequential steps in the iterations was taken $10^{-5}$ eV units as convergence criterion. The convergence for the Hellmann-Feynman forces per unit cell was taken to be $10^{-4}$ eV/Å and Gaussian smearing of 0.05 eV was used. Integration over the Brillouin zone was done with the equivalent of a 24×24×1 Monkhorst-Pack k-point grid for a single graphene unit cell[17]. Supercells of 5×5 unit cells were used to calculate the energy barriers with a vacuum layer of 15 Å between periodic images in the vertical direction.

Protons and Li ions were modeled using different approaches. Protons were modelled using the charged hydrogen pseudopotential provided by the Vienna ab-initio Simulation Package (VASP). On the other hand, due to the lack of pseudopotentials for Li-ions in the VASP, these were modelled using the pseudopotential of a neutral Li atom. *A posteriori*, we found that a Li atom transfers ~1 $e$ of charge to the 2D material several angstroms away from the ring structures. Hence, for small distances to film, the Li atom effectively behaves as a Li$^+$ ion, which allows for the estimate of the energy barrier of Li$^+$ ion using this pseudopotential – see below.

In the calculations, three different carbon-atom ring structures were considered: defect-free graphene, graphene with Stone-Wales (SW) defect and graphene with a 5-8-5 defect (see Supplementary Fig 6 a). These correspond to 6-, 7- and 8- carbon-atom ring structures, respectively. A proton or a Li atom were placed at a distance $z$ in the perpendicular direction to the center of these ring-structures; the system was allowed to fully relax and the energy of the atom plus film was calculated as a function of $z$. This calculation was repeated for different $z$ values in steps of 0.1 Å. The energy barrier of the transport process was determined by the largest energy step in this trajectory.

The energy barrier for proton penetration through the defect structures decreased notably with the number of atoms in the rings (Supplementary Figure 6b). For Stone-Wales and 5-8-5 defects, the barriers were 0.88 eV and 0.78 eV, respectively. These values are ~1.5 and ~2 times lower than the barrier found for defect-free graphene (1.39 eV), respectively. The nature of the barrier was also different amongst the ring structures. Unlike defect-free graphene and Stone-Wales defect, protons bind to 5-8-5 defects. Dissociating from these latter defects leads to the largest energy step in their trajectory across the film (Supplementary Figure 6b).

Our calculations for Li atoms revealed that these bind to graphene several angstroms (>4 Å) away from the 2D film because the Li atom donates about one electron to graphene. This was the case irrespective of whether or not a defect was present. For this reason, we find that at distances ≲3 Å from the film, the Li atom interacts with the 2D films like a Li$^+$ ion. The energy barrier can be estimated from the largest step in the trajectory for distances ≲3 Å from the film (Supplementary Figure 6c,d). Within this approximation, we found that the energy barrier for Li ion penetration through the ring-structures decreased notably with the number of atoms in the rings. The energy barriers posed by Stone-Wales defects and 8-atom rings were 3.6 eV and 1.5 eV, respectively. These are ~2 and ~3 times lower than for 6-atom rings (7.54 eV), respectively.

This analysis also explains the experimentally observed Li/H selectivity in disordered graphene. Because $\sigma$ depends exponentially on the energy barrier posed by the different ring-structures, transport through 8-atom rings dominate the response. Our calculations reveal that the barrier posed by 8-atom rings to Li ion penetration is about twice higher than for protons, which explains the observed Li/H selectivity in disordered graphene.

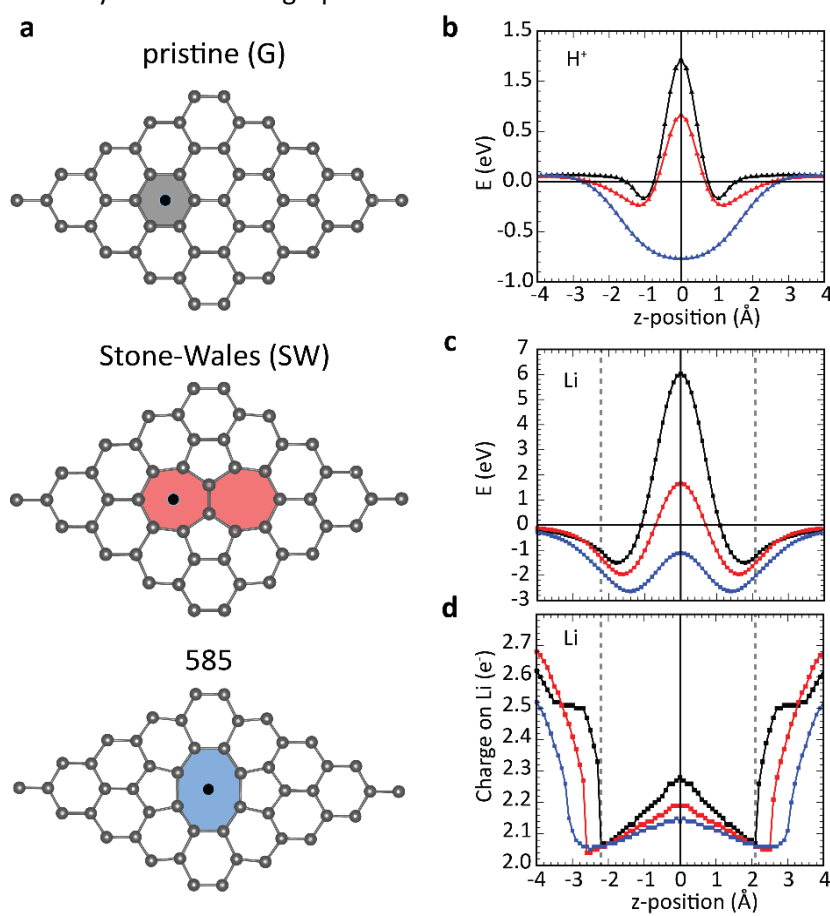

**Supplementary Figure 6 | Density functional theory calculations**. **a**, Schematic of the different supercells studied. The ring-structures of interest are marked in colors (grey, pristine graphene; red, Stone-Wales; blue, 585). The penetration site for protons and Li are marked with a black dot. **b**, Energy profiles for a proton translocating through graphene (grey), SW (red) and 5-8-5 (blue) as a function of distance to the center of the ring-structures (*z*-position). **c,** Energy profiles for a Li atom translocating through graphene (grey), SW (red) and 5-8-5 (blue) as a function of distance to the center of the ring structures (z-distance). Grey dotted lines mark the *z*-position range within which the Li atom interacts with the 2D film as a Li-ion. **d**, Electron charge localized on the Li atom as function of *z*-position. Grey dotted lines mark the *z*-position range within which the electron charge on the (initially neutral) Li-atom drops to ~2 *e*. Given the +3 total charge in Li nuclei, this yields a Li$^+$ ion.